# Optically controllable magnetism in atomically thin semiconductors

Kai Hao[1*], Robert Shreiner[1,2*], & Alexander A. High[1,3†]

[1]Pritzker School of Molecular Engineering, University of Chicago, Chicago, IL 60637, USA.

[2]Department of Physics, University of Chicago, Chicago, IL 60637, USA.

[3]Center for Molecular Engineering and Materials Science Division, Argonne National Laboratory, Lemont, IL 60439, USA.

[*]These authors contributed equally to this work.

[†]To whom correspondence should be addressed: ahigh@uchicago.edu

**Electronic states in two-dimensional (2D) layered materials can exhibit a remarkable variety of correlated phases including Wigner-crystals, Mott insulators, charge density waves, and superconductivity[1–5]. Recent experimental[6,7] and theoretical[8–12] research has indicated that ferromagnetic phases can exist in electronically-doped transition metal dichalcogenide (TMD) semiconductors, but a stable magnetic state at zero magnetic field has eluded detection. Here, we experimentally demonstrate that mesoscopic ferromagnetic order can be generated and controlled by local optical pumping in monolayer tungsten diselenide (WSe$_2$) at zero applied magnetic field. In a spatially resolved pump-probe experiment, we use polarization-resolved reflectivity from excitonic states as a probe of charge-carrier spin polarization. When the sample is electron-doped at density $n_e \sim 10^{12}$ cm$^{-2}$, we observe that a local, circularly-polarized, microwatt-power optical pump breaks the symmetry between equivalent ferromagnetic spin configurations and creates magnetic order which extends over mesoscopic regions as large as 8 μm x 5 μm, bounded by sample edges and folds in the 2D semiconductor. The experimental signature of magnetic order is

**circular dichroism (CD) in reflectivity from the excitonic states, with magnitude exceeding 20% at resonant wavelengths. The helicity of the pump determines the orientation of the magnetic state, which can be aligned along the two principle out-of-plane axes. In contrast to previous studies in 2D materials that have required non-local, slowly varying magnetic fields to manipulate magnetic phases, the demonstrated capability to control long-range magnetism and corresponding strong CD with local and tunable optical pumps is highly versatile. This discovery will unlock new TMD-based spin and optical technologies and enable sophisticated control of correlated electron phases in two-dimensional electron gases (2DEGs).**

Due to favorable material properties and tuning capabilities, TMDs are a rapidly emerging platform for the study and manipulation of collective phases in 2DEGs. For free electrons in TMDs, the combination of a large effective mass ~0.44 $m_e$ and reduced dielectric screening creates a Bohr-radius that is only slightly larger than the lattice constant[7,13]. As a result, the energy of Coulombic interactions can be appreciably larger than energies associated with phase-space filling, leading to collective ordering of electronic states dictated by long-range exchange interactions[5,8–12,14]. In particular, exchange interactions are predicted to create a variety of spin- and valley-polarized magnetic phases at low carrier densities $n$ ~ $10^{12}$ - $10^{13}$ cm$^{-2}$ [8,10,11]. These carrier densities are readily achievable in TMDs by electrostatic doping[6,7,15–17].

Recently, experiments have shown that under applied magnetic field and in certain doping regimes, electrons in molybdenum disulphide ($MoS_2$) and molybdenum diselenide ($MoSe_2$) exhibit magnetic order with near-complete spin-polarization far beyond the predictions of a simple thermal population model[6,7,15]. The spin polarization manifests as circular dichroism in reflectivity and photoluminescence measurements of the excitonic states, and was initially attributed to an interaction-enhanced electronic g-factor (so-called giant paramagnetism)[15] or the emergence of ferromagnetic order[6,7]. In the ferromagnetism model, the spin polarization is due to strong exchange interactions which favor the formation of a spin-polarized state in both the K and K' valleys[6,11]. Follow-up experiments demonstrated that the system transitions from a ferromagnetic to a paramagnetic phase with increased doping, suggesting direct electronic control over the electron-electron interactions and correlated phases[7]. These studies present compelling evidence that the magnetic ordering at moderate carrier densities is ferromagnetic in nature and that spin-polarized electronic states should persist even at zero magnetic field. However, while ferromagnetic order could be expected at zero applied magnetic field, no net magnetization or spin polarization was observed. This absence was attributed to the lack of a global symmetry breaking mechanism[7].

Optical pumping is a conceivable mechanism to break the symmetry between equivalent spin configurations in TMDs. Recent studies have shown that pumping individual monolayers or heterostructures of TMDs with circularly polarized light can generate spin imbalances with microsecond-long relaxation times[16,18–20]. For $WSe_2$ monolayers in the electron doped region, which is the main focus of this work, resident electrons can be dynamically spin/valley-polarized by continuous pumping with circular light[21]. Photo-generated electrons excited in a selected valley by the circularly polarized pump will preferentially relax to the opposite valley due to fast spin-conserving intervalley scattering. Additionally, the intravalley recombination of resident electrons with photo-generated holes forming dark excitons can enhance the asymmetry of the valley populations[20]. The resulting spin-polarization is maintained in the presence of the continuous pump as these processes occur on timescales faster than the spin relaxation rate[20,21].

Moreover, due to the relatively low free charge carrier densities $n \approx 10^{12} - 10^{13}$ cm$^{-2}$, a significant population of resident carriers may be spin-polarized, potentially sufficient to break the symmetry between ground-state spin configurations and stabilize the ferromagnetic order.

Here, we study the impact of above-bandgap, circularly polarized optical pumping on h-BN encapsulated monolayers of WSe$_2$. The heterostructure layout and optical image of the sample D1 are presented in Figures 1a and 1b. The doping level in the monolayer can be controlled by applying gate voltage between the few-layer graphene contact and top gate and manifests in the appearance of neutral and charged excitonic resonances in the reflection spectra, Figure 1c. We first focus on low temperature measurements at T = 4 K within the moderately electron-doped region, where singlet (X$^-$$_S$) and triplet (X$^-$$_T$) trions are clearly observed[22–24]. Reflection from a circularly polarized supercontinuum laser provides a probe of the local, valley-selective optical response. In the absence of pumping, balanced reflection of σ+ and σ- polarized light is observed (Fig. 1d), indicating that the probe laser does not generate any symmetry breaking. Next, we pump the sample with a 660 nm diffraction-limited continuous wave laser with a spot diameter of 500 nm and a power of 7.8 µW. To demonstrate the nonlocality of pump-induced effects and to eliminate the influence of photoluminescence in detection, the probe spot is separated by nearly 8 µm from the pump (Fig. 1b). The reflection spectra under σ+ polarized pumping are markedly different – the triplet (singlet) trion dominates the probe signal co(cross)-polarized to the pump (Fig. 1e). This pump-induced circular dichroism is characterized by $CD = \Delta R^+ - \Delta R^-$, where $\Delta R^{+,-} = (R^{+,-}_{on} / R^{+,-}_{off}) - 1$ is the differential reflectivity comparing the σ+,- probed reflection in the presence ($R^{+,-}_{on}$) and absence ($R^{+,-}_{off}$) of the pump. The CD signal displays amplitudes approaching 10% and inverts with the sign of the pump polarization (Fig. 1f).

The CD is a direct signature of electron spin/valley-polarization, in which singlet and triplet trions are formed preferably in the opposite valleys (see Fig. 1e inset). To quantify the spin polarization, we determine the valley-dependent oscillator strengths of the trion states by fitting the reflection contrast with a Breit-Wigner-Fano lineshape (see Supplementary Information). Since the valley-dependent charge density correlates with the oscillator strength of the transition[25,26], we estimate that 90%(10%) of charges reside in the valley cross(co)-polarized with the pump even at 8 μm pump-probe separation. This corresponds to a spin polarization $P_s$ = 0.77, where $P_s = (A^+ - A^-) / (A^+ + A^-)$ and $A^{+,-}$ is the probe-polarization-selective oscillator strength of the trion state under optical pumping. For an electron doping density of $n \approx 1.8 \times 10^{12}$ cm$^{-2}$ (see Supplementary Information), this yields a spin population imbalance of ~$1.4 \times 10^4$ μm$^{-2}$. Small disparities in the local optical environment and sample inhomogeneity may subtly modify the relationship between the CD signal and the underlying spin polarization, however, this extracted value provides a reasonable approximation.

Remarkably, the optical pump generates a near-complete spin polarization that persists micrometers away from the pump location. We next study this spatial dependence in more detail. Figure 2a shows a photoluminescence (PL) map of the region of interest (ROI) of the monolayer flake. The central dark area corresponds to a bilayer region. Keeping the pump at the fixed point x = 0 μm (star in Fig. 2a), the probe is scanned along the length of the monolayer (dashed line in Fig. 2a). Strong CD signal is detected, reaching an amplitude of 20% (Fig. 2b). Moreover, the CD signal shows significant energetic variation, indicating that the spin polarization is robust to sample inhomogeneity. These observations imply the maintenance of long-range spin polarization built up under circularly polarized optical pumping.

To further characterize the spatial profile of the spin polarization, we map the CD signal across the entirety of the ROI. Figure 2c depicts the CD associated with the singlet trion peak as the probe is scanned across the flake while the σ+ pump remains fixed. CD is clearly observed within the pristine portion of the ROI, except for in the bilayer region where no resonance peak nor CD signal are found. When the sign of the pump polarization is flipped to σ- (Fig. 2d), the CD signal inverts everywhere. To investigate the correlation between the pump-probe displacement and the CD intensity, we plot a cross section of the CD amplitude around the ROI (Fig. 2e). No clear CD signal decay is observed as the separation increases. Indeed, in the c to b direction, we observe an increase in CD signal. The spatial inhomogeneity of the sample is depicted in Figure 2f. The trion resonance energy varies by up to 30 meV within the ROI, while the CD signal is still robust. However, more prominent imperfections apparently destroy the spin polarization. The purple and blue dashed lines indicate wrinkles and residue in the heterostructure observed under microscope imaging, which correspond to observable dips in photoluminescence (Fig. 2a). The CD signal terminates upon crossing the noted defects.

To gain additional insight into the origin of the mesoscopic CD, we vary the doping concentration. As depicted in Figure 1c, we can access the intrinsic, hole-doped and highly electron-doped region by varying the gate voltage. Here, we study the long-range CD (*i.e.*, pump-probe separation of 8 µm, Fig. 1b) within these different doping regions. As shown in Figure 3a, no CD signal is observed in the intrinsic region, implying the optically induced CD is correlated with free carriers in the system. As in Figure 1f, we observe strong CD co(cross)-polarized to the pump from the triplet (singlet) features. At higher doping concentrations, while the heavily-doped charged exciton $X^{--}$ [27] (Mahan exciton[7] or attractive polaron[15]) is clearly observed in reflectivity, there is no observable CD. We further characterize this by plotting the extracted peak amplitude (Fig. 3b) and CD amplitude (Fig. 3c) against the estimated doping density for different species of excitonic states. We observe that at 1.75 V, $n_c \approx 4 \times 10^{12}$ cm$^{-2}$, the system transitions from observable singlet and triplet trions with strong CD, to X$^{--}$ exciton states with no observable CD. Therefore, our measurements clearly correlate the CD with the singlet and triplet trion spectral features in the electron-doped regime. We also observe CD signal in the hole-doped region. We note that in the hole-doped regime, we are unable to fully quench the exciton reflectivity, indicating limited doping capability.

We also investigate the strength of the CD with respect to changes in temperature (on a second sample D2), pump polarization, and excitation power. Figure 4a depicts the doping dependent CD spectra at different temperatures. The CD signal vanishes at T = 30 K, even though reflection spectra still exhibit clear resonance features of singlet/triplet trions and X$^{--}$ states (see Supplementary Information). The temperature dependent triplet CD amplitude and estimated spin polarization are plotted in Figure 4b, showing a rapid transition from an unpolarized ($P_s = 0.06$) to a polarized ($P_s = 0.87$) spin state as the temperature goes below T = 15 K. To gauge the impact of pump polarization on CD, we sweep the polarization from σ+ to linear and then σ- (Fig. 4c), observing the singlet CD signal increase from its minimum value through zero to its maximum value. When scanning in the opposite direction, the CD signal shows the same trend, vanishing at linear pump polarization. Lastly, we measure the impact of pump power on CD (Fig. 4d). The pump induces mesoscopic CD even with powers as low as 100 nW, and the CD signal begins saturating at ~2 µW. No hysteresis is observed when the pump power is scanned in the opposite direction, similar to the pump polarization scan. While the magnitude of the CD varies with pump power, the spatial profile of the CD remains nearly uniform within an order of magnitude variation in pumping power (Fig. 4e).

We attribute the mesoscopic spin polarization to ferromagnetic order in the TMD induced by the optical pump. This claim is supported by the following observations: (I) In our measurements, mesoscopic spin-polarization emerges within the same carrier density regime as theoretically predicted and experimentally studied ferromagnetic phases in TMDs. In theoretical models of electron-electron interactions in TMDs[8–12,14], exchange inter- and intra-valley coupling lead to spin-polarized ferromagnetic phases at electron densities around $n_e \sim 10^{12}$ cm$^{-2}$. These models were validated by experimental studies of magnetic phases in electron-doped monolayer molybdenum disulphide (MoS$_2$) in an external magnetic field. In that case, strong exchange intervalley interactions – compared to the small spin-orbit splitting of the conduction bands in MoS$_2$ – lead to band inversion and the spin polarization of resident electrons across both K and K' valleys (*i.e.*, spin-polarized, but not valley-polarized electrons)[6,11]. In contrast, WSe$_2$ monolayers exhibit an order of magnitude larger spin-orbit splitting in the conduction band[10]. Consequently, the predicted ferromagnetic ground state consists of spin/valley-polarized resident electrons[10], in agreement with our circular dichroism results. (II) The temperature dependence of the CD signal, which represents the magnetization[28,29], displays a trend qualitatively consistent with other 2D ferromagnetic materials[26,27], and criticality fits indicate a Curie temperature $T_C$ = 15 K with a critical exponent of 0.113, close to the value of 0.125 for a 2D Ising model[30] (see Supplementary Information). (III) The disappearance of spin polarization with increased electron doping signifies a 1$^{st}$-order phase transition from a ferromagnetic to a paramagnetic phase. This is in full agreement with previous experimental[7] and theoretical studies[8–12,14] of magnetic phase transitions in TMDs. (IV) We additionally observe signatures of long-range order in the hole-doped regime, consistent with theory[8,10], although yet to be shown experimentally.

While spin diffusion and giant-paramagnetic order are alternative mechanisms for generating long-range spin polarization, the experimental data does not support these processes. There is no obvious correlation between the CD amplitude and the separation of the pump and probe, in stark contrast to a diffusion process[17]. Additionally, the rapid and complete disappearance of long-range CD with relatively small changes in carrier density is not readily explained by simple diffusion. Furthermore, we do not apply a magnetic field, eliminating the possibility of interaction-enhanced paramagnetic order[15]. The temperature dependence also is incompatible with a paramagnetic phase, failing to maintain a 1/T trend at low temperature. In total, our experimental observations of mesoscopic spin polarization are fully consistent with the emergence and control of ferromagnetic order by optical pumping.

A primary innovation in our study is the use of a continuous-wave optical pump to break the symmetry between equivalent magnetic phases and stabilize mesoscopic magnetic order against fluctuations. Magnetic state fluctuations explain both the absence of spin order below $T_c$ without the optical pump and the lack of observable hysteresis in our time-averaged measurements. The fluctuations may arise due to nano-ampere-scale leakage currents in our sample, which can destabilize the magnetic state by injecting unpolarized electrons. The optical pump then stabilizes a single magnetic state against fluctuations by selectively valley-pumping spin-polarized electrons, thereby breaking the symmetry between degenerate magnetic states and preferentially favoring the formation of a co-polarized magnetic state. This mechanism is fundamentally different from previously reported all-optical control of magnetism, which is based on heating and inverse Faraday effects[31–34]. We also emphasize that the optical pump should not be considered directly analogous to an external magnetic field as a symmetry breaking mechanism. In contrast to Zeeman interaction in a magnetic field, the optical pump does not change the energetics of the system. Additionally, the dynamics of the pump-induced spin polarization may lead to time-dependent evolution of the magnetic phase, which is not captured by our steady state measurements.

Critically, while the optical pump acts locally, the magnetic order is stabilized mesoscopically, extending well beyond the sub-micron pumping region to the boundaries of the monolayer. Our measurements afford some insight into the nature of this non-locality. We observe that, rather than spreading from the pump location in a non-linear fashion, the magnetism emerges uniformly over the sample with increasing pump power. This suggests that in the absence of optical pumping, long-range magnetic phases exist in the sample, but the phases fluctuate temporally between spin configurations. The optical pump acts to pin one of the two long-range magnetic phases against fluctuations. The micron-scale domain size in our measurements is comparable in size to magnetic domains in other 2D magnetic materials[29,35]. This picture contrasts with a model in which the electronic ground state at zero applied magnetic field consists of local, nanoscale magnetic domains[7] – in this picture, the pumping would not be expected to generate mesoscopic polarization beyond the locally controlled domains. Follow up studies will fully elucidate the dynamics of the magnetic phases and the process of long-range stabilization.

We present evidence for optically controllable ferromagnetic order in TMDs. The recent discovery of magnetic 2D materials has generated significant excitement due to their novel integration and heterostructure possibilities[36–38]. Our research clearly establishes TMDs as a 2D magnetic material, albeit with very different properties than more conventional 2D magnets – critically, the magnetic configuration can be fully tuned non-locally with optical fields and electronic gating. Moreover, the local optical pump stabilizes the magnetic state, even at low sub-microwatt power, providing finer spatial resolution for the study and control of magnetic domain structure. These unique features open new avenues for probing the previously inaccessible physics of magnetic order in two-dimensional semiconductors, prompting future experimental and theoretical investigations of the temporal dynamics and spatial formation of domains under optical pumping. Additionally, TMDs are a prototypical platform for explorations of correlated phenomena in 2DEGs, and we show that optical pumps provide a powerful tool for understanding and controlling these systems. For instance, magnetic phases and their circular dichroism could be utilized to manipulate and probe Mott insulators and Wigner crystals[1–3].

Furthermore, our findings will accelerate technological developments utilizing TMDs, already a leading material platform for investigating next-generation spin-, valley-, and optoelectronics[39–41]. Specifically, the discovery of optically-reconfigurable magnetism and circular dichroism in atomically thin semiconductors will stimulate the design of non-reciprocal optoelectronics and photonics[42,43], such as on-chip all-optical isolators with built-in optical memory. Lastly, our research creates a bridge between magnetism and optical control in TMDs, which can be leveraged for direct interfacing between integrated photonics and magnetic solid-state memories[44], suggesting new routes for neuromorphic optical computing[45].

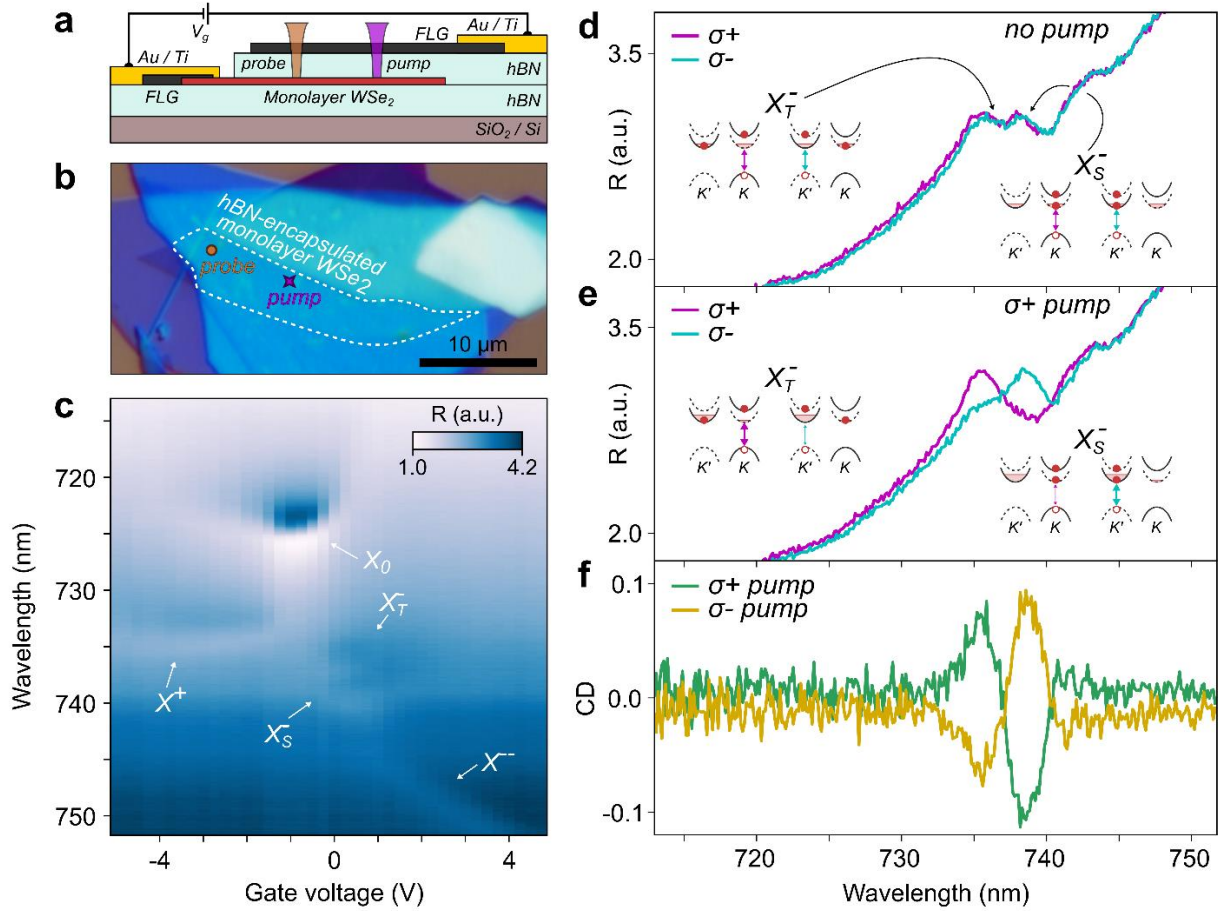

**Figure 1: Sample under study. a,** Schematic of hBN-encapsulated WSe$_2$ monolayer with few-layer graphene top gate and contacts. The optical pump and probe are spatially separated. **b,** Optical microscope image of sample D1. **c,** Gate dependent reflection spectra of the WSe$_2$ sample. The excitonic resonance features are labeled correspondingly. **d,** σ+ and σ- reflection spectra at 0.5 V, where the singlet and triplet trion features are well resolved. Inset: Singlet and triplet trion configurations showing balanced valley populations. **e,** σ+ and σ- reflection spectra at 0.5 V under σ+ pumping. Insert: Schematic of singlet and triplet trion in optically pumped spin/valley-polarized electron bath. **f,** Circular dichroism (CD) spectra under σ+ and σ- pumping. T = 4 K and pump power is 7.8 μW.

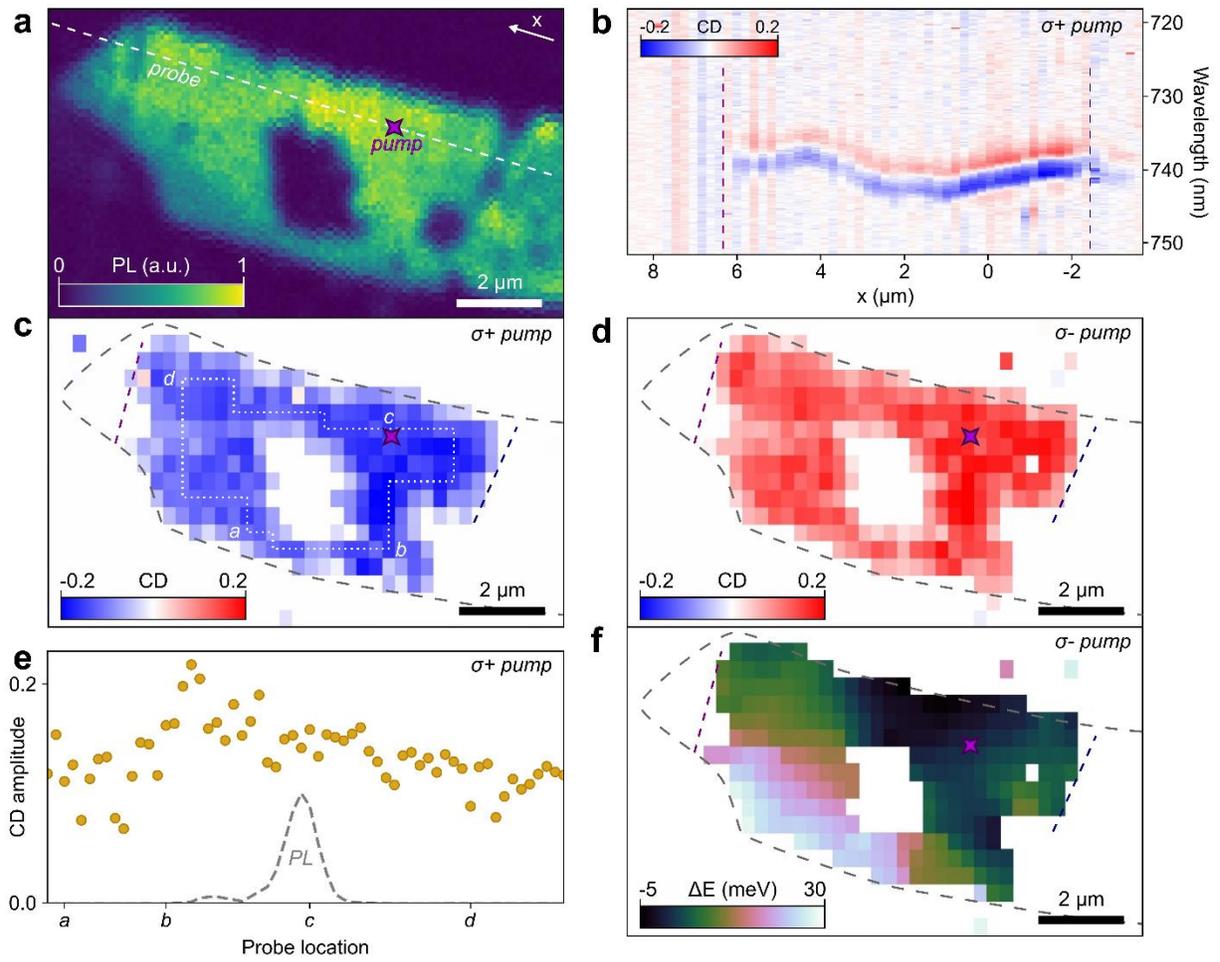

**Figure 2: Spatial profile of the spin polarization. a,** Photoluminescence (PL) map of the ROI. **b,** Spatially dependent CD spectra under σ+ pumping. The fixed pump location x = 0 µm is given by the star in (**a**), while the probe is scanned along the dashed white line in (**a**). **c,d,** Map of the CD amplitude across the whole ROI under σ+(**c**) and σ-(**d**) pumping. **e,** CD amplitude along the dashed contour in (**c**). Dashed gray line depicts the PL intensity for fixed-position pumping. **f,** Map of peak energy shifts of the singlet CD signal compared to the value at the pump location. Purple and blue dashed lines correspond to wrinkles on the sample. T = 4 K and pump power is 7.8 µW.

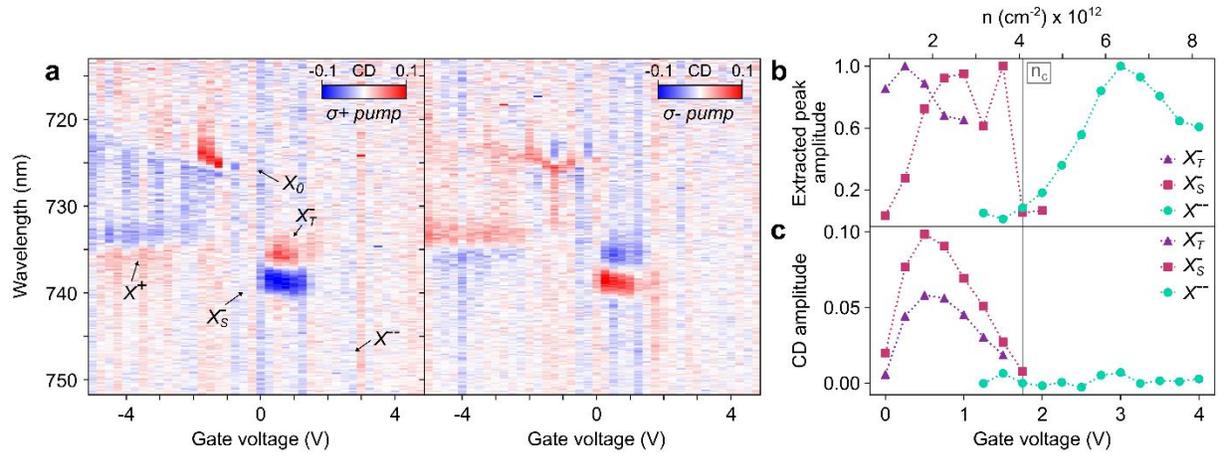

**Figure 3: Gate dependence of CD spectra. a,** Gate dependent CD spectra probed 8 µm from the pump (Fig. 1b) under σ+ (left) and σ- (right) pumping. Excitonic states are labeled corresponding to the features in reflection spectra. **b,** Extracted peak amplitude versus gate voltage (doping level). **c,** CD amplitude versus gate voltage (doping level). The critical electron density $n_c$ is indicated. T = 4 K and pump power is 7.8 µW.

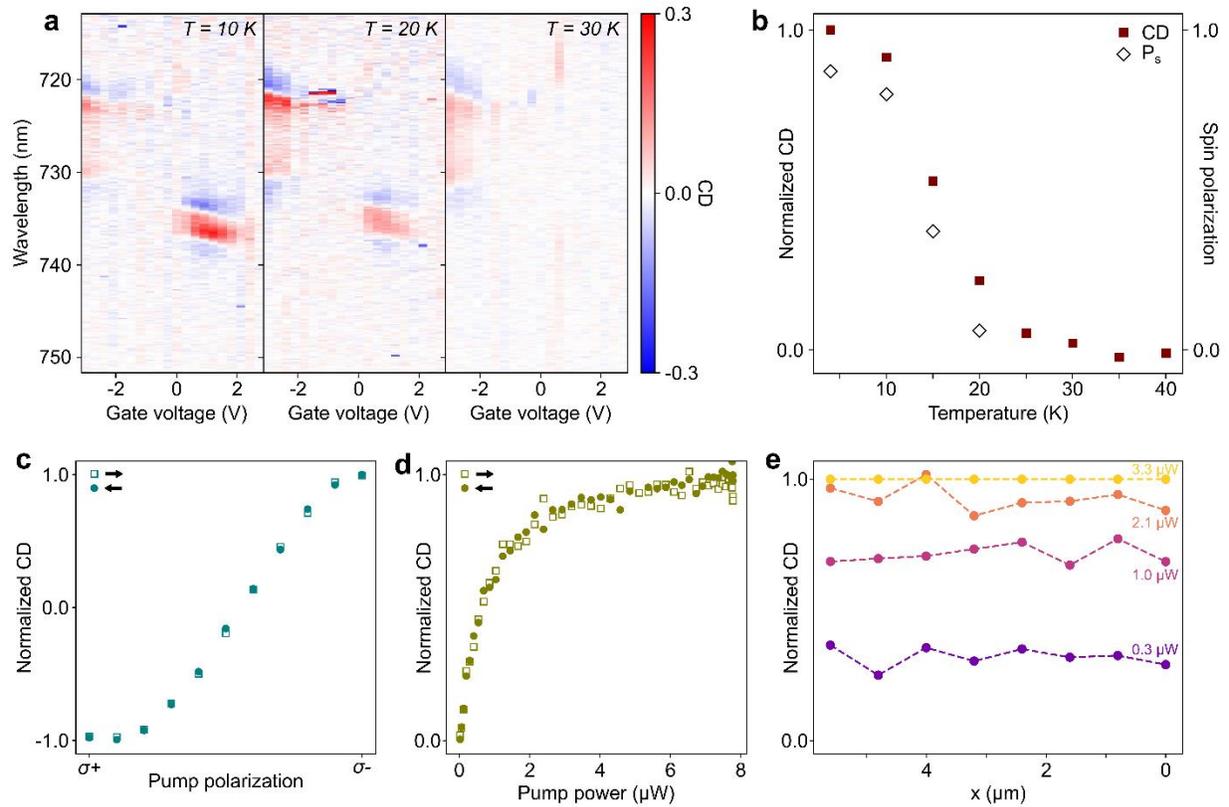

**Figure 4: Stability of CD signal. a,** Gate-dependent CD spectra of sample D2 at 10 K, 20 K and 30 K (pump-probe offset of 2.2 µm, pump power of 10 µW). **b,** Temperature dependence of triplet CD amplitude (red squares) selected from gate-dependent CD spectra at gate voltage of 1.2 V. Corresponding spin polarizations (hollow diamond) are extracted from reflection spectra. **c,** Polarization dependence of singlet CD amplitude of sample D1 (pump-probe offset of 1.6 µm, pump power of 7.8 µW). Hollow squares (solid circles) correspond to sweeping the polarization from σ+ to σ- (σ- to σ+). **d,** Power dependence of singlet CD amplitude of sample D1 (pump-probe offset of 1.6 µm). Hollow squares (solid circles) correspond to increasing (decreasing) pump power. **e,** Spatial dependence of singlet CD amplitude of sample D1 under selected pump powers, normalized to the 3.3 µW profile. T = 4 K unless otherwise stated.

**Acknowledgements:** This work made use of the Pritzker Nanofabrication Facility part of the Pritzker School of Molecular Engineering at the University of Chicago, which receives support from Soft and Hybrid Nanotechnology Experimental (SHyNE) Resource (NSF ECCS-2025633), a node of the National Science Foundation's National Nanotechnology Coordinated Infrastructure. This work also made use of the shared facilities at the University of Chicago Materials Research Science and Engineering Center, supported by the National Science Foundation under award number DMR-2011854. Funding was provided by Army Research Office Grant #W911NF-20-1-0217. The authors acknowledge Andrew Kindseth, You Zhou, and Jiwoong Park for helpful discussions.

**Author contributions:** A.H. conceived the study. K.H. & R.S. performed sample fabrication, experiments, and analyses. All authors contributed to writing the manuscript.

**Competing interests:** The authors declare no competing financial interests.

**Supplementary Information** is available for this paper.

**Correspondence** to Alexander A. High.

**Data availability:** The data that support the findings of this study are available from the corresponding author on reasonable request.

# Optically controllable magnetism in atomically thin semiconductors

## Supplementary Information

Kai Hao[1*], Robert Shreiner[1,2*], & Alexander A. High[1,3†]

[1]Pritzker School of Molecular Engineering, University of Chicago, Chicago, IL 60637, USA.

[2]Department of Physics, University of Chicago, Chicago, IL 60637, USA.

[3]Center for Molecular Engineering and Materials Science Division, Argonne National Laboratory, Lemont, IL 60439, USA.

[*]These authors contributed equally to this work.

[†]To whom correspondence should be addressed: ahigh@uchicago.edu

**Sample fabrication**

The monolayer tungsten diselenide ($WSe_2$), hexagonal boron nitride (hBN) and few-layer graphene (FLG) flakes are mechanically exfoliated from commercial bulk crystal ($WSe_2$ – 2D Semiconductor; hBN and FLG - HQGraphene) onto $Si/SiO_2$ chips. The thickness and cleanliness of the flakes are first examined with optical microscopy and then by atomic force microscopy (AFM) (Fig. S1a). We used an all-dry transfer method[1] to fabricate the hBN-encapsulated $WSe_2$ stack with FLG top gate and contact. Electrodes are patterned via photolithography, and then deposited by e-beam physical deposition with 5 nm Ti and 95 nm Au (Fig. S1b). We also note that sample D2 has a layer of PMMA on top of the completed heterostructure.

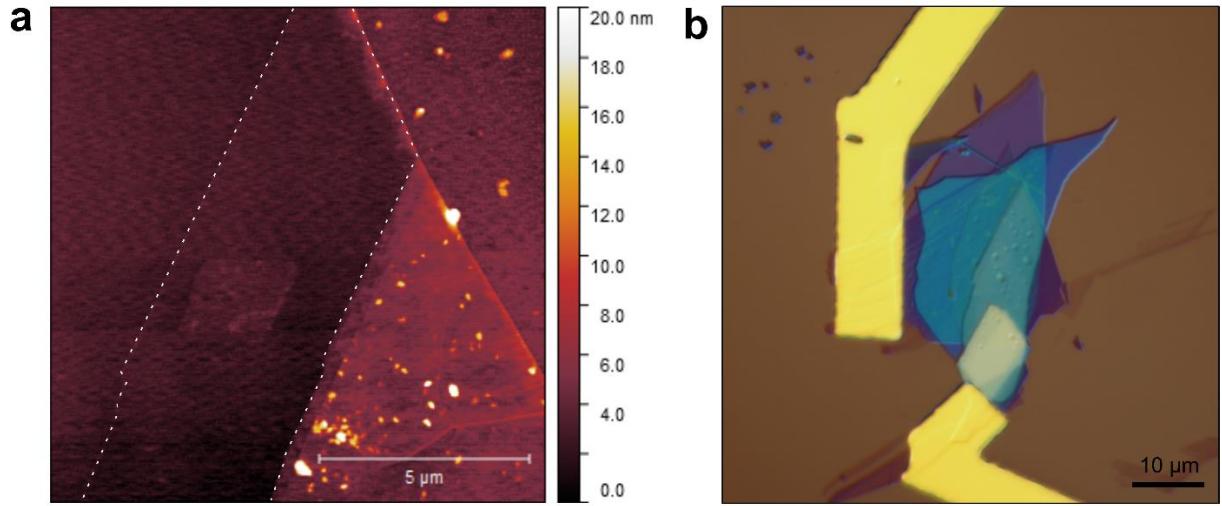

**Figure S1: Sample fabrication. a,** AFM image of the WSe$_2$ flake on Si/SiO$_2$ chip with monolayer region outlined. **b,** Optical microscope image of the full heterostructure D1 with patterned contacts.

**Optical measurement setup**

The samples were kept in a close loop cryostat (Montana Instruments) at 4 K during the experiment unless otherwise claimed. The optical setup is depicted in Fig. S2. The two galvo mirrors control the pump and probe beam independently to realize spatial scans. We use a 660 nm diode laser (Thorlabs) as the optical pump exciting through port 3 with a band pass filter to spectrally clean the pump laser. A supercontinuum laser (YSL Photonics) is deployed as the broadband probe sent through port 1. The reflection of the supercontinuum from the sample is collected by port 2 and fiber coupled to a spectrometer with CCD (Teledyne Princeton Instruments) to realize spectrally resolved reflection measurements. The polarizations of each beam are independently controlled by linear polarizers and half-wave plates to allow different pump/probe polarization combinations.

To map the photoluminescence (PL) from the sample (Fig. 2a), we use ports 1 and 2 as the pump and collection channels, respectively. The same 660nm diode laser is used for pumping. The PL signal is collected from port 2 and detected by an APD (Excelitas). By scanning galvo mirror 1, the co-localized pump and collection are simultaneously moved across the sample, realizing the PL mapping.

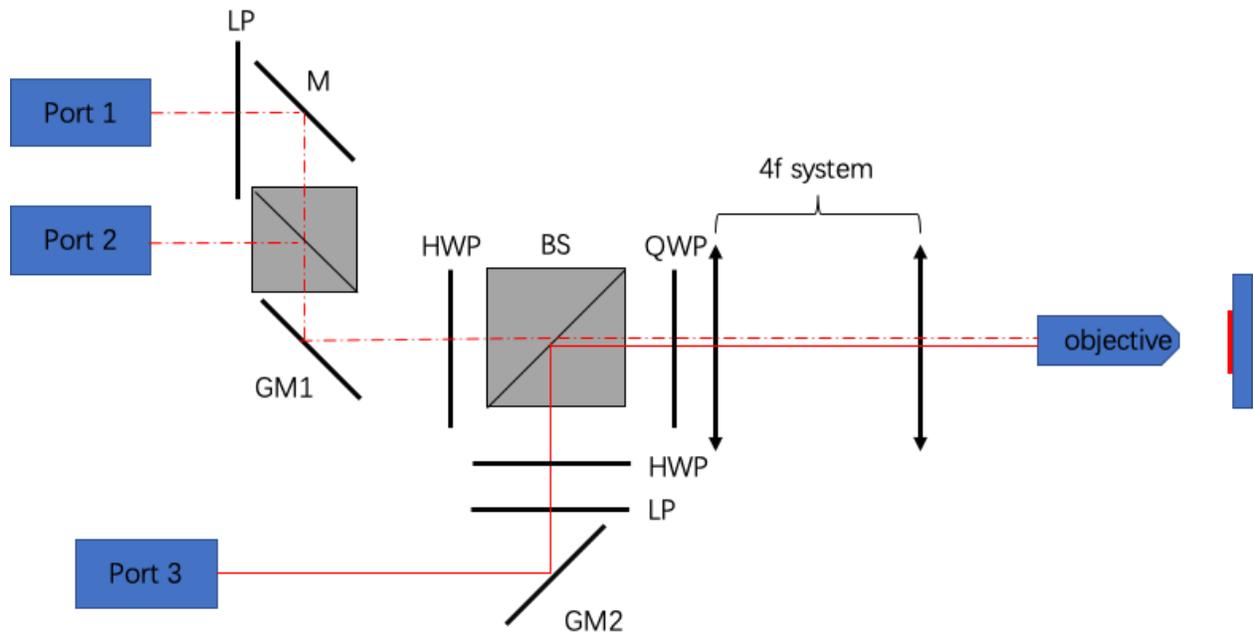

**Figure S2: Optical setup.** Setup components are: Port – fiber launcher, M – mirror, GM – Galvo mirror, LP – linear polarizer, HWP – half-wave plate, QWP – quarter-wave plate, BS – beam splitter, 4f system – lens pair, objective – NA = 0.75.

**Capacitor model of charge density estimation**

To estimate the charge carrier density under electrostatic gating, we model the heterostructure as a parallel plate capacitor [2]: the FLG and monolayer $WSe_2$ flakes are the electrodes separated by the dielectric top hBN flake. The capacitance per unit area is given by $C = \frac{\epsilon_{hBN} \epsilon_0}{d_{hBN}}$. Using the relative permittivity of hBN $\epsilon_{hBN} = 3.76$ [1] and the hBN thickness $d_{hBN} = 11.5$ nm, we find $C \approx 290$ nF cm$^{-2}$. The carrier density as a function of gate voltage is given by $n(V_g) = C (V_g - V_0)$, where $V_0$ is the gate voltage corresponding to the intrinsic regime with no carrier doping. By comparison with the neutral exciton feature in reflection (Fig. 1c), we approximate $V_0 = -0.5$ V.

**Line shape analysis**

To estimate the valley/spin polarization of the free carriers, we extract the total oscillator strength of the trions in the two valleys. By comparing the oscillator strengths, we approximate the relative density of states in each valley, which give an estimation of the polarization of the free carriers[3,4].

The reflection contrast corresponding to the two trion features from Fig. 1e is depicted in Fig. S3 (dots), which shows an asymmetric Fano-like line shape. The reflection contrast is given by $RC = \frac{R}{R_{ref}} - 1$, where R is the bare reflectivity (Fig. 1e) and $R_{ref}$ is a reference reflectivity where the trion oscillator strengths disappear, under large gate bias or in the intrinsic region[5]. Such spectra can be captured by a Breit-Wigner-Fano (BWF) line shape fitting:

$$I(\omega) = I_0 \frac{(1 + \frac{\omega - \omega_0}{q\Gamma})^2}{1 + (\frac{\omega - \omega_0}{\Gamma})^2}$$

where $q$ is the parameter which captures the asymmetry of the line shape. When $\frac{1}{q} \to 0$, the function converges to the Lorentzian line shape, where $I_0$ is the amplitude, $\omega_0$ is the center energy, and $\Gamma$ is the linewidth[6].

By fitting the RC to a superposition of BWF line shapes with two resonances corresponding to the singlet and triplet trions (Fig. S3 lines), we extract the fitting parameters shown in table T1. The oscillator strength of each resonance is proportional to the area of the Lorentzian line shape $A = I_0 \Gamma$. By comparing the relative oscillator strengths for each state under co- and cross-circular pumping, we can estimate the ratio of the density of states in each valley. We take this ratio, 90:10 for the triplet trion, as the ratio between the valley-polarized spin states of the resident electrons.

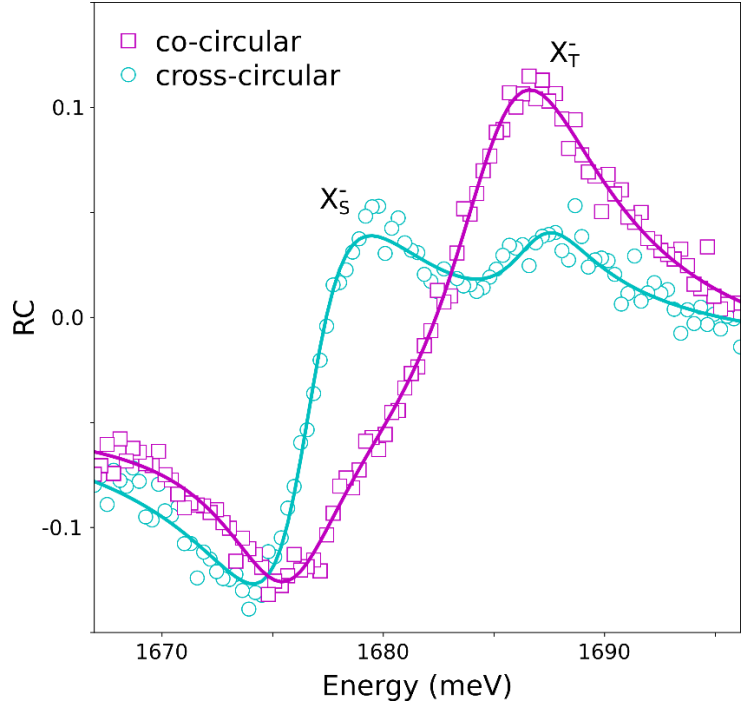

**Figure S3: Line shape fitting.** Polarized reflection contrast spectra showing singlet ($X_S^-$) and triplet ($X_T^-$) trion features under optical pumping. The pink (blue) dots correspond to the probe being polarized co- (cross-) circular to the pump. BWF fittings are plotted.

**Table T1. Fitting the trion features under optical pumping.**

| Triplet | $I_0$ | $\omega_0$ (meV) | $\Gamma$ (meV) | $A = I_0\Gamma$ | Ratio | $P_s$ |
|---|---|---|---|---|---|---|
| Co-circular | 0.151 | 1685.4 | 4.1 | 0.62 | ~90:10 | 0.77 |
| Cross-circular | 0.035 | 1687.0 | 2.3 | 0.08 | | |
| **Singlet** | $I_0$ | $\omega_0$ (meV) | $\Gamma$ (meV) | $A = I_0\Gamma$ | Ratio | $P_s$ |
| Co-circular | 0.0005 | 1675.8 | 3.3 | 0.0016 | ~99:1 | -0.98 |
| Cross-circular | 0.075 | 1676.5 | 2.7 | 0.20 | | |

**Sample D2 characterization and temperature dependence**

The temperature-dependent measurements (Fig. 4a,b) are performed on a second sample (D2). Figure S4a shows gate-dependent reflection spectra at 4 K with a pump-probe separation of 2.2 µm, which display similar doping regimes as sample 1 (Fig. 1c). In the electron-doped region, strong CD signal is observed. Figure S4b depicts the reflection spectra at 30 K. While the electron-doped CD vanishes at this elevated temperature (Fig. 4a), the singlet and triplet trion features remain, confirming that the temperature dependence of the CD signal probes that of the spin polarization. Temperature dependent data was also taken with overlapped pump and probe. The CD amplitude shows a very similar trend as that with a 2.2 µm pump-probe offset (Fig. S4c). The criticality fit[7] $\alpha \, (T_c - T)^\beta$ is applied to the CD data, extracting a critical temperature

$T_c = 15.0\ K\ (15.3\ K)$ and a critical exponent $\beta = 0.113\ (0.106)$ for the offset (overlapping) pump-probe configuration. For the 2D Ising model, the expected critical exponent is $\beta = 0.125$.

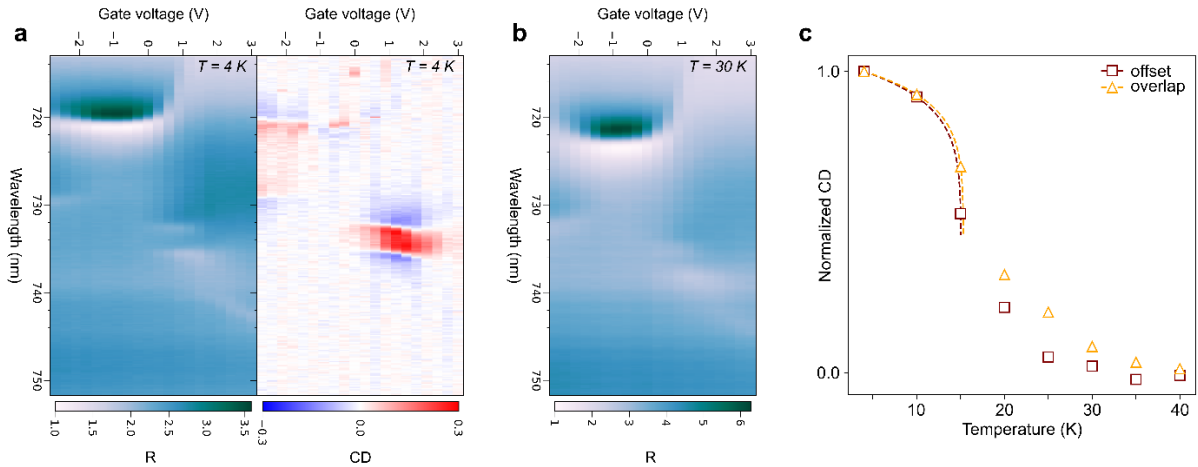

**Figure S4: Sample 2 characterization. a,** Gate-dependent reflection (left) and CD spectra (right) at 4 K with a pump-probe separation of 2.2 µm. **b,** Reflection spectra at 30 K. **c,** Temperature dependence of triplet CD amplitude selected from gate-dependent CD spectra at gate voltage of 1.2 V. Hollow squares (triangles) correspond to experimental data for offset (overlapping) pump and probe with respective fittings (dashed lines).